\documentclass[twocolumn,showpacs,preprintnumbers,amsmath,amssymb,aps]{revtex4}
\usepackage[dvips]{graphicx}% Include figure files
\usepackage{dcolumn}% Align table columns on decimal point
\usepackage{bm}% bold math
\def\be{\begin{equation}}
\def\ee{\end{equation}}

\usepackage{color}
\def\col{\textcolor{black}}

\def\bea{\begin{eqnarray}}
\def\eea{\end{eqnarray}}
\usepackage{amsfonts}
\usepackage{amssymb}
\begin{document}

\title{An Alternative Interpretation for the Moduli Fields of the Cosmology
Associated to Type IIB Supergravity
with Fluxes}
\author{Tonatiuh~Matos*$^{1,2}$, Jos\'e-Rub\'en Lu\'evano$^3$, Hugo
Garc\'{\i}a-Compe\'an$^{2,4}$ and J. Alberto
  V\'azquez\footnote{Part of the Instituto Avanzado de
    Cosmolog\'ia (IAC) collaboration http://www.iac.edu.mx/}$^2$}

\affiliation{$^{1}$CIAR Cosmology and Gravity Program, Department of Physics
and Astronomy,\\
University of British Columbia, Vancouver, British Columbia, Canada, V6T
1Z1}

\affiliation{$^{2}$ Departamento de F{\'\i}sica, Centro de Investigaci\'on y
de Estudios Avanzados del IPN,
A.P. 14-740, 07000 M\'exico D.F., M\'exico.}

\affiliation{$^{3}$Departamento de Ciencias B\'asicas, Universidad
Aut\'onoma Metropolitana-Azcapotzalco, C.P. 02200 M\'exico, D.F.,
M\'exico.}

\affiliation{$^{4}$ Centro de Investigaci\'on y de Estudios Avanzados del
IPN, Unidad Monterrey,
Cerro de las Mitras 2565, Col. Obispado, cp. 64060, Monterrey N.L.,
M\'exico}

\begin{abstract}
The aim of this work is to provide a basis to interpret the
dilaton as the dark matter of the universe, in the context of a
particular cosmological model derived from type IIB supergravity
theory with fluxes. In this theory, the dilaton is usually
interpreted as a Quintessence field. But, with this alternative
interpretation we find that (in this supergravity model) the model
gives a similar evolution and structure formation of the universe
compared with the $\Lambda$CDM model in the linear regime of
fluctuations of the structure formation. Some free parameters of
the theory are fixed using the present cosmological observations.
In the non-linear regime there are some differences between the
type IIB supergravity theory with the traditional CDM paradigm.
The supergravity theory predicts the formation of galaxies earlier
than the CDM and there is no density cusp in the center of
galaxies. These differences can distinguish both models and might
give a distinctive feature to the phenomenology of the cosmology
coming from superstring theory with fluxes.
\end{abstract}

\date{\today}


\maketitle
% relativity and gravitation
% cosmology

\section{Introduction}

One of the main problems in physics now is to know the nature of
the dark matter and the understanding of the accelerated expansion
of the universe.
%-------------------------------------------------------------------
These two phenomena have been observed in the last years and now
there are a number of observations supporting the existence of the
dark matter \cite{DM} and the accelerated expansion of the
universe as well \cite{DE}.
%----------------------------------------------------------------------------
On the other side, one of the main problems of superstring theory
is that there is not a real phenomenology which can support the
theory. Usually, superstring theory is supported only by its
mathematical and internal consistency, but not by real experiments
or observations. For some people, like the authors, one way of how
superstring theory can make contact with phenomenology is through
the cosmology \cite{CdeSS}. In the last years, a number of new
observations have given rise to a new cosmology and to a new
perception of the universe (see for example \cite{rewC}). In
superstring theory there are 6 extra dimensions forming a compact
internal Calabi-Yau manifold \cite{GSW}.
%-----------------------------------------------------------------------------
Size and shape of this manifold manifests, at the four dimensional
low energy effective field theory, a series of scalar fields
(moduli of the theory) many of which apparently have not been seen
in nature.
%------------------------------------------------------------------------------------
In particular, two fields, the {\it dilaton} and the {\it axion},
are two very important components of the theory which can not be
easily fixed.
%------------------------------------------------------------------------------------------
In fact, one should find a physical interpretation for these
fields or give an explanation of why we are not able to see them
in nature.
%-----------------------------------------------------------------------------------
One interpretation is that there exist a mechanism for eliminating
these fields during the evolution of the universe \cite{DamourSS}.
Recently, one of the most popular interpretations for the dilaton
field is that it can be the dark energy of the universe, $i.e.$ a
Quintessence field \cite{QuintdeSS}. These last interpretations
have been possible because after a non trivial compactification,
the dilaton field acquires an effective potential. This effective
potential makes possible to compare the dilaton field with some
other kinds of matter \cite{QuintdeSS}.
%--------------------------------------------------------------------
In this work we are giving the dilaton a {\it different}
interpretation supposing that it is the dark matter instead of the
dark energy \cite{Cho}.
%-----------------------------------------------------------------
Such attempts have been carried over in the past with other
dilaton potentials \cite{dilatonDM}. Here, we will be very
specific starting with an effective potential derived recently
from the type IIB supergravity theory. The main goal of this work
is to show that this interpretation could be closer to a realistic
cosmology as the interpretation that the dilaton is the dark
energy. We will see that the late cosmology is very similar to the
$\Lambda$CDM one with this alternative interpretation.
Nevertheless, we will also see that it is necessary to do
something else in order to recover a realistic cosmology from
superstring theory.
%------------------------------------------------------------------------------------
On the other hand, a great deal of work has been done recently, in
the context of string compactifications with three-form fluxes
(R-R and NS-NS) on the internal six-dimensional space and the
exploration on their consequences in the stabilization of the
moduli fields including the dilaton $\Phi$ and axion $C$
\cite{fluxes}. Moduli stabilization has been used also in string
cosmology to fix other moduli fields than the volume modulus
including dilaton+axion and Kahler moduli \cite{branden}. For a
description of more realistic scenarios, see \cite{krause}.

In the context of the type IIB supergravity theory on the ${\bf
T}^6/\mathbb{Z}_2$ orientifold with a self-dual three-form fluxes,
it has been shown that after compactifaying the effective
dilaton-axion potential is given by \cite{Frey02}
\bea
V_{dil}&=& \frac{M_P^4}{4(8\pi)^3}h^2\,e^{-2\Sigma_i\sigma_i}\,
\left[ e^{-\Phi^{(0)}}\cosh \left(\Phi-\Phi^{(0)}
  \right)  \right. \nonumber
  \\
  &+&\left.\frac{1}{2}\,e^{\Phi}(C-C^{(0)})^{2}-e^{-\Phi^{(0)}}\right],
  \label{eq:Poteff}
\eea
where $h^2= {1 \over 6} h_{mnp}h_{qrs} \delta^{mq} \delta^{nr}
\delta^{ps}$. Here $h_{mnp}$ are the NS-NS integral fluxes, the
superscript $(0)$ in the fields stands for the fields in the
vacuum configuration and finally $\sigma_i$ with $i=1,2,3$ are the
overall size of each factor ${\bf T}^2$ of the ${\bf
T}^6/\mathbb{Z}_2$ orientifold (in \cite{Frey02} there is a
mistprint in the potential \ref{eq:Poteff}). Here we will simplify
the system supposing that the moduli fields $\sigma_i$ are
constant for the late universe.

For the sake of simplicity in the derivation of the potential
(\ref{eq:Poteff}), some assumptions were made \cite{Frey02}. One
of them is the assumption that the tensions of D-branes and
orientifold planes cancel with the energy $V_{dil}$ at
$\Phi=\Phi_0$ and $C = C_0$. An assumption on initial conditions
is that the dilaton is taken to deviate from equilibrium value,
while the complex structure moduli are not.
It is also assumed that
(\ref{eq:Poteff}) has a global minimum $\Phi_0$, such that
$V(\Phi_0)=0$. Also that, the complex moduli are fixed and only
the radial modulus $\sigma$ feels a potential when the
dilaton-axion system is excited. These assumptions make the model
more simple, but still with the sufficient structure to be of
interest in cosmological and astrophysical problems.

In order to study the cosmology of this model, it is convenient to
define the following quantities
$\lambda\sqrt{\kappa}\phi=\Phi-\Phi^{(0)}$,
$V_0=\frac{M_P^4}{4(8\pi)^3}h^2\,e^{-2\Sigma_i\sigma_i}\,e^{-\Phi^{(0)}}$,
$C-C^{(0)}=\sqrt{\kappa}\psi$  and $\psi_0=e^{\Phi^{(0)}}$, where
$\lambda$ is the string coupling $\lambda =
e^{\langle\Phi\rangle}$ and $\lambda\sqrt{2\kappa}$ is the reduced
Plank mass $M_p/\sqrt{8\pi}$. With this new variables, the dilaton
potential transforms into
\bea V_{dil}&=& {V_0}\, \left( \cosh \left( \lambda\sqrt{\kappa}\phi
  \right) -1\right)+\frac{1}{2}V_0\,{e^{{\lambda} \sqrt{\kappa}\phi
}}{{\psi_0}}^{2}\kappa\psi^{2}\nonumber\\
&=&V_{\phi}+e^{{\lambda} \sqrt{\kappa}\phi }V_{\psi}. \label{eq:V}
\eea

In some works the scalar field potential (\ref{eq:V}) is suggested
to be the dark energy of the universe, that means, a Quintessence
field \cite{QuintdeSS}\cite{Frey02}.
%--------------------------------------------------------------
In this work we are not following this interpretation to the
dilaton field.
%---------------------------------------------------------------
Instead of this, we will interpret the term ${V_0}\, \left(
\cosh\left(\lambda\sqrt{\kappa}\phi\right) -1\right)$ as the dark
matter of the universe\cite{coshDM}, \cite{tmatosDM}. The
remaining term in $V_{dil}$ contains the contribution of the axion
field $C$.
%---------------------------------------------------------------------------
This is what makes the difference between our work and
previous ones.
%---------------------------------------------------------------
This interpretation allows us to compare the cosmology derived
from the potential (\ref{eq:V}) with the $\Lambda$ cold dark
matter ($\Lambda$CDM) model.
%-----------------------------------------------------------------------
The rest of the fields coming from superstrings theory can be
modeled as usual, assuming that this part of the matter is a
perfect fluid. This perfect fluid has two epochs: radiation and
matter dominated ones.
%-------------------------------------------------------------------
In order to consider both epochs we write the matter component as
matter and radiation, with a state equation given by
${\dot\rho_b}+3\,H\,\rho_b=0$ and
${\dot\rho_{rad}}+4\,H\,\rho_{rad}=0$. For modelling the dark
energy we can take the most general form supposing that it is also
a perfect fluid with the equation of state given by
${\dot\rho_L}+3\,\gamma_{DE}\,H\,\rho_{L}=0$, where $\gamma_{DE}$
is smaller than 1/3 and can even be negative in the case it
represents a phantom energy \cite{phantom} field. It is just zero
if $\rho_L$ represents the cosmological constant $L=\rho_L$.

\section{The Cosmology}
Now, we proceed to describe the different epochs of the universe
using this new interpretation. We can easily distinguish two
behaviors of the scalar field potential: the exponential and the
power laws.
%--------------------------------------------------------
In the early universe the exponential behavior dominates the scalar fields
potential. In this case we have the following analysis.

\noindent {\it Inflation.-} In this epoch, the scalar field
potential can be written as
\begin{equation}
V= {V_0}\, \exp \left( \lambda\sqrt{\kappa}\phi\right)\left(
1 +\frac{1}{2}\,\kappa  {{\psi_0}}^{2}\psi^{2}\right),
\end{equation}
because the exponential dominates completely the scenario of the
evolution of the dilaton potential. The distinctive feature during
this period is that the presence of the fluxes generate a
quadratic term in the Friedman equation. The scalar field density
$\rho=1/2\,{\dot\phi}^2+1/2\,{\dot\psi}^2\,e^{\tilde\lambda\sqrt{\kappa}\phi}+V$
appears quadratic in the field equations,
\be
H^2=\frac{\kappa}{3}\,\rho\left(1+\frac{\rho}{\rho_0}\right).
\ee
Under this conditions it is known that these potentials are always
inflationary in the presence of these fluxes \cite{lidsey}.
Nevertheless, exponential potentials are inflationary without
branes, in the traditional Friedman cosmology, only if
$\lambda^2<2$ (see for example \cite{liddleBook}). Therefore, if
we suppose that $\lambda^2>2$, the dilaton potential (\ref{eq:V})
is not inflationary without the quadratic density term. Thus, as
the universe inflates, the quadratic term becomes much more
smaller than the linear term and we recover the Friedman equation
$H^2=\frac{\kappa}{3}\,\rho$, where the exponential potential is
not inflationary anymore.
%----------------------------------------------------
For these values of $\lambda$ this gives a natural graceful-exit
to this scalar field potential \cite{Lidsey:2001nj}. It remains to
study which is the influence of the axion potential to this epoch
\cite{erandy}.

\begin{figure}[htb]
\begin{center}
\includegraphics[width=9cm]{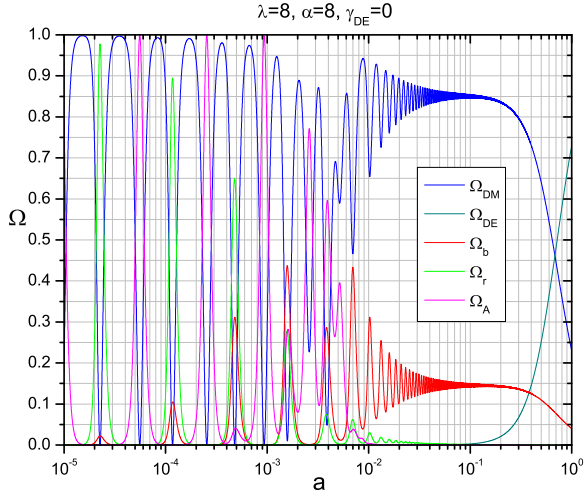}
\caption{\label{fig:completo1} Plot of the dynamics of the
$\Omega$'s in the type IIB superstring theory with fluxes. Observe
how this theory predicts a similar behavior of the matter content of
the universe as the $\Lambda$CDM model. Here, the initial values of
the dynamical variables at redshift $a=1$ are: $x=0,\, A=0,\,
u=\sqrt{0.23},\, v=1000,\, \Omega_{DE}=0.7299,\,\Omega_{b}= 0.04,\,
\Omega_{rad}=4\times 10^{-5}$, and $w$ is determined by the Friedman
restriction. The values for the constants are $\alpha = \lambda =
8,\,\tilde\lambda= 7,\,\Psi_0\sqrt{V_0/\rho_L}=5000$. In all
figures, the integration was made using the Adams-Badsforth-Moulton
algorithm (variable step size). Each curve contains over $8\times
10^{5}$ points.}
\end{center}
\end{figure}

\begin{figure}[htb]
\begin{center}
\includegraphics[width=9cm]{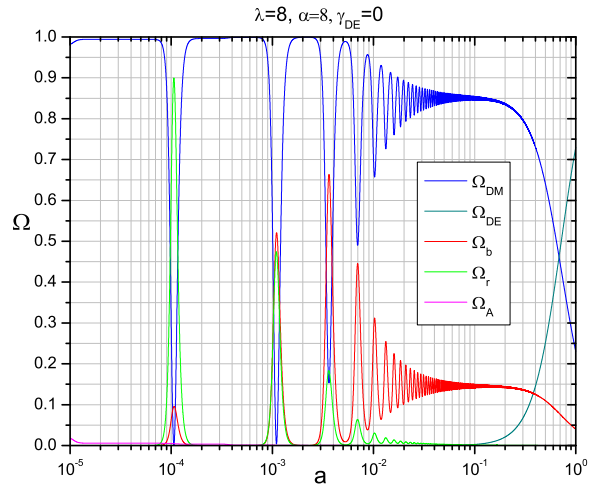}
\caption{\label{fig:completo2} Plot of the dynamics of the
$\Omega$'s in the type IIB superstring theory with fluxes. Initial
values of the dynamical variables at redshift $a=1$ are the same as
in Fig \ref{fig:completo1}. The values for the constants are
$\Psi_0\sqrt{V_0/\rho_L}=5000,\,\alpha=8,\,\lambda=8,\,\tilde\lambda=0$.
Each curve contains over $8\times 10^{5}$ points.}
\end{center}
\end{figure}

\noindent{\it Densities Evolution.-} The evolution of the
densities is quite sensible to the initial conditions. Let us
study the example of evolution shown in Fig.\ref{fig:completo1}.
As in the $\Lambda$CDM model, here also the recombination period
starts around the redshift $10^{3}$. The first difference we find
between $\Lambda$CDM model and the IIB superstrings theory is just
between the redshifts $10^{3}$ and $10^{2}$, where the interaction
between the dilaton and matter gives rise to oscillations of the
densities. It is just in this epoch where we have to look for
observations that can distinguish between these two models. In
this epoch the scalar field is already small
$\lambda\sqrt{\kappa}|\phi|<4$ and approaches the minimum of the
potential in $\phi=0$. Thus, potential (\ref{eq:V}) starts to
behave as a power low potential, simulating a type $\phi^2$ field.
Therefore it is not surprising that this potential mimics very
well the dark matter behavior. In the $\Lambda$CDM model, dark
matter is modeled as dust and it is well known that power low
potentials mimics dust fluids as they oscillate around the minimum
of the potential \cite{kolb}. In figures Fig.\ref{fig:completo1}
and Fig.\ref{fig:completo2} this behavior is confirmed.

Nevertheless, for redshifts bigger than $1/a-1=z\sim10^3$, there
are remarkable differences between the superstring model and the
CDM one. The interaction of the dilaton field with matter provokes
to be very difficult that radiation dominates the universe, thus
big bang nucleosynthesis never takes place, at lest in a similar
way as in the CDM paradigm. Let us explain this point. The dilaton
field interacts with matter through the factor
$e^{\tilde\alpha(\Phi-\Phi^{(0)})}\,F^2=e^{\alpha\sqrt{\kappa}\phi}\,F^2$,
being $F$ the field strength of the matter contents. Thus,
Lagrangian for the superstrings system is
\begin{equation}
  \mathcal{L} =\sqrt{-g} \left( R - \mathcal{L}_\phi -
e^{\tilde\lambda\sqrt{\kappa} \phi}
  \mathcal{L}_\psi - e^{\alpha\sqrt{\kappa} \phi} \mathcal{L}_{matter}
\right) \, ,
  \label{eq:ss_Lagrangian}
\end{equation}
where we have differentiated the scalar field potential coupling
constant $\lambda$ from the axion-dilaton coupling constant
$\tilde\lambda$ in order to generalized and clarify the cosmology
of the system. In (\ref{eq:Poteff}) both are the same
$\lambda=\tilde\lambda$. The individual Lagrangians for the
dilaton and axion fields respectively are,

\begin{equation}
    \mathcal{L}_\phi = \frac{1}{2} \partial^\sigma \phi
    \partial_\sigma \phi + V_\phi \, , \quad \mathcal{L}_\psi =
    \frac{1}{2} \partial^\sigma \psi
    \partial_\sigma \psi + V_\psi \, .
    \label{eq:ss_Lagrangians}
\end{equation}

\begin{figure}[htb]
\begin{center}
\includegraphics[width=9cm]{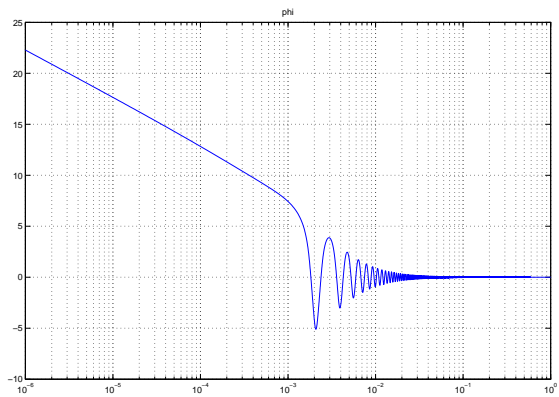}
\caption{\label{fig:phi} Plot of the behavior of the scalar field
$\phi$. The scalar field starts from big values and riches very
fast its minimum where it starts to oscillate. We plot
$\lambda\sqrt{\kappa}\phi$, for $\lambda =20$.}
\end{center}
\end{figure}

Thus, in a flat Friedman-Robertson-Walker space-time the
cosmological field equations are given by
\begin{eqnarray}
H^{2}&=&\frac{\kappa}{3}\,\left( \frac{1}{2}{\dot\phi}^{2}+ \frac
{1}{2}{\dot\psi}^{2}e^{\col{\tilde\lambda}\sqrt{\kappa}\phi}+V_{\phi}+e^{\col{\tilde\lambda}
\sqrt{\kappa}\phi
}V_{\psi} \right. \nonumber\\
&+& \left.  \left({\rho_{b}}
+\rho_{rad}\right)e^{\alpha\sqrt{\kappa}\phi} + {\rho_{L}}
\right),  \label{eq:Friedman}\\
{\ddot \phi} &+&3\,H {\dot \phi} +
\frac{d\,\col{V_\phi}}{d\phi}
=\col{\tilde\lambda\sqrt{\kappa}}\,e^{\col{\tilde\lambda}\sqrt{\kappa}\phi}\left(\frac{1}{2}\col{\dot\psi^2}
-\col{V_\psi}\right) \nonumber\\
&-& \left.
\alpha\sqrt{\kappa}\,e^{\alpha\sqrt{\kappa}\phi}\col{(\rho_b+\rho_r)}
\right. \label{eq:phi}
\\
{\ddot \psi} &+& 3\,H {\dot \psi} +
\frac{d\,V_\psi}{d\psi}=-\col{\tilde\lambda
\sqrt{\kappa}}\,\dot\phi\,\dot\psi, \label{eq:psi}
\\
{\dot\rho_b}&+&3\,H\,\rho_b=0, \label{eq:b}
\\
{\dot\rho_{rad}}&+&4\,H\,\rho_{rad}=0, \label{eq:rad}
\\
{\dot\rho_L}&+&3\,\gamma_{DE}\,H\,\rho_{L}=0, \label{eq:DE}
\end{eqnarray}
where the dot stands for the derivative with respect to the
cosmological time and $H$ is the Hubble parameter $H={\dot a}/a$.
In order to analyze the behavior of this cosmology, we transform
equations (\ref{eq:Friedman})-(\ref{eq:DE}) using new variables
defined by
\bea
x&=&\frac{\sqrt{\kappa}}{\sqrt{6}}\frac{{\dot\phi}}{H},\,\,
A=\frac{\sqrt{\kappa}}{\sqrt{6}}\frac{{\dot\psi}}{H}e^{\frac{1}{2}\col{\tilde\lambda}\sqrt{\kappa}\phi},
\label{eq:x1}\\
y&=&\frac{\sqrt{\kappa}}{\sqrt{3}}\frac{\sqrt{\rho_b}}{H}e^{\frac{1}{2}\alpha\sqrt{\kappa}\phi},\,\,
z=\frac{\sqrt{\kappa}}{\sqrt{3}}\frac{\sqrt{\rho_{rad}}}{H}e^{\frac{1}{2}\alpha\sqrt{\kappa}\phi},
\label{eq:z1}\\
u&=&\frac{\sqrt{\kappa}}{\sqrt{3}}\frac{{\col{\sqrt{V_\phi}}}}{H},\,\,
v=\frac{\sqrt{\kappa}}{\sqrt{3}}\frac{{\sqrt{V_2}}}{H},
\label{eq:u}\\
l&=&\frac{\sqrt{\kappa}}{\sqrt{3}}\frac{\sqrt{\rho_L}}{H},\,\,
w=\col{\frac{\sqrt{\kappa}}{\sqrt{3}}\frac{{\sqrt{V_\psi}}}{H}e^{\frac{1}{2}\tilde\lambda\sqrt{\kappa}\phi}},
\label{eq:New_var_def} \eea
where we have used the definition of the potentials
$V_\phi=2\,V_0\sinh(1/2\,{\sqrt{\kappa}\lambda\phi})^2$,
$V_2=2\,V_0\cosh(1/2\,{\sqrt{\kappa}\lambda\phi})^2$ and
$\col{V_\psi=\frac{1}{2}V_0\kappa{\psi_0}^2\psi^2}$ such that
$\col{V=V_\phi+V_\psi e^{\tilde\lambda\sqrt{\kappa}\phi}}$ is the
total scalar field potential. With these definitions equations
(\ref{eq:Friedman})-(\ref{eq:DE}) transform into
\begin{eqnarray}
&x^{\prime }&=-3\,x-\sqrt {\frac{3}{2}}\, \left( \lambda uv+
\alpha(y^2+z^2)+\tilde\lambda (w^2-{A}^{2}) \right)+\frac{3}{2}
\Pi\,x,\nonumber
\\\label{eq:x} \\
&A^{\prime } &=-3\,A-\sqrt
{3}\,\frac{\psi_0\sqrt{V_0}}{\sqrt{\rho_L}}w\,l-\sqrt{\frac{3}{2}}\col{\tilde\lambda}\,A\,x
+\frac{3}{2}\Pi\,A, \label{eq:A}
\\
&y^{\prime } &=\frac{3}{2}\, \left( \Pi-1+
\alpha\sqrt{\frac{2}{3}}x\right) y, \label{qe:y}
\\
&z^{\prime } &=\frac{3}{2}\, \left( \Pi-\frac{4}{3}+
\alpha\sqrt{\frac{2}{3}}x \right)\,z, \label{eq:z}
\\
&u^{\prime } &=\sqrt {\frac{3}{2}}\lambda\,v\,x+\frac{3}{2}
\Pi\,u, \label{eq:u}
\\
&v^{\prime } &=\sqrt {\frac{3}{2}}\lambda\,u\,x+\frac{3}{2}
\Pi\,v, \label{eq:uv}
\\
&l^{\prime } &=\frac{3}{2}\, \left( \Pi-{\gamma_{DE}} \right)\,l,
\label{eq:l}
\\
&w^{\prime } &=\sqrt
{3}\,\frac{\psi_0\sqrt{V_0}}{\sqrt{\rho_L}}\,A\,l
+\tilde\lambda\sqrt{\frac{3}{2}}w\,x+\frac{3}{2}\,\Pi\,w
\label{eq:w}
\end{eqnarray}
where now prime stands for the derivative with respect to the
$N$-foldings parameter $N=\ln(a)$. The quantity $\Pi$ is defined
as
\begin{equation}
\Pi= 2\,{x}^{2}+2\,{A}^{2}+{y}^{2}+\frac{4}{3}\,{z}^{
2}
\end{equation}
The Friedman equation (\ref{eq:Friedman}) becomes a constriction
of the variables such that
\be x^2+A^2+y^2+z^2+u^2+l^2+w^2=1.
\label{eq:constriccion} \ee
The density rate quantities $\Omega_x=\rho_x/\rho_{critic}$ can be
obtained using the variables (\ref{eq:x1}) - (\ref{eq:New_var_def}), one arrives
at
\begin{eqnarray}
\Omega_{DM}&=& x^2+u^2, \nonumber\\
\Omega_{DE}&=& l^2, \nonumber\\
\Omega_{b}&=& y^2, \nonumber\\
\Omega_{rad}&=& z^2, \nonumber\\
\Omega_{A}&=& 1-x^2-u^2-y^2-z^2-l^2, \label{eq:densidades}
\end{eqnarray}
where $\Omega_{DM},\,\Omega_{DE},\,\Omega_{b},\,\Omega_{rad}$ and
$\Omega_{A}$ respectively are the density rates for the dark
matter (dilaton field), dark energy (cosmological constant),
baryons, radiation and axion field. For the definition of this
last one we have used the constriction (\ref{eq:constriccion}).
Equations (\ref{eq:x})-(\ref{eq:w}) are now a dynamical system.
The complete analysis of this system will be given elsewhere
\cite{zorro}, but the main results are the following. $1.-$ The
system contains many critical points, some of them are atractors
with dark matter dominance, other with dark energy dominance.
$2.-$ The system depends strongly on the initial conditions. One
example of the evolution of the densities is plotted in
Fig.\ref{fig:completo1} and Fig.\ref{fig:completo2}, where we show
that the densities behave in a very similar way as the
corresponding ones of the $\Lambda$CDM model before redshifts
$10^{2}$, which seem to be a generic behavior. The free constants
$\lambda$, $\alpha$ and $\tilde\lambda$ are given in each figure.
On the other side, we can see that after redshifts $z\sim 10^{3}$
one finds that $|\phi|\,<\,0.04\,m_{Planck}$ and oscillating goes
to zero, such that its exponential is bounded
$0.01\,<\,e^{-\lambda\sqrt{\kappa}\phi}\,<\,1$ see Fig.
\ref{fig:phi}. In other words, it takes the exponential more than
13 Giga years to change from 0.01 to 1. %%

\begin{figure}[htb]
\begin{center}
\includegraphics[width=9cm]{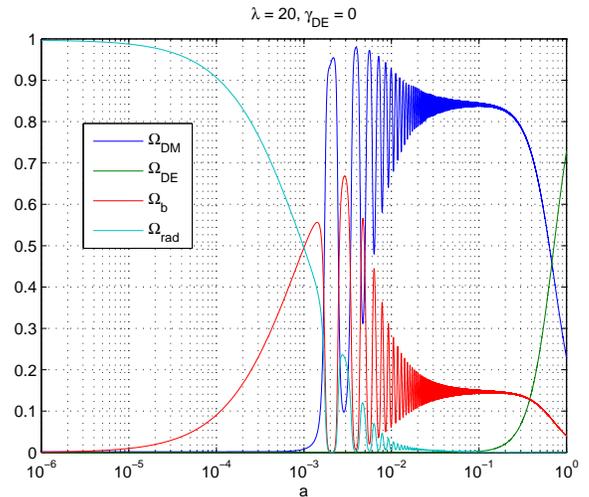}
\caption{\label{fig:pegado} Plot of the dynamics of the $\Omega$'s
in the type IIB superstring theory with fluxes. Observe how this
theory predicts a similar behavior of the matter content of the
universe as the $\Lambda$CDM model, even for redshifts beyond
$10^3$. Here radiation dominates the universe for values less than
$a\sim10^{-3}$ and big bang nucleosynthesis takes place as in the
CDM model. Here $\lambda=\alpha=20$, the initial values of the
dynamical variables at redshift $a=1$ are: $x=0,\, A=0,\,
u=\sqrt{0.23},\, v=1000,\, \Omega_{DE}=0.7299,\,\Omega_{b}=
0.04,\, \Omega_{rad}=4\times 10^{-5}$, and $w$ is determined by
the Friedman restriction. Each curve contains over $3\times
10^{5}$ points.}
\end{center}
\end{figure}

However, there is one fact that takes our attention in
Fig\ref{fig:completo1}. We see from the behavior of the densities
on the early universe after redshifts $\sim10^3$, that radiation
does not dominates the rest of the densities as it is required for
big bang nucleosynthesis. This fact can also be seen as follows.
In a radiation dominated universe we might set $l=y=u=v=w=A=0$, in
that case we can see by inspection of (\ref{eq:x})-(\ref{eq:w})
that there is no way that radiation remains as a dominant
component of the system. The situations radically change if we put
$\tilde\lambda=0$ in system (\ref{eq:x})-(\ref{eq:w}), in this
case radiation has no problems to be dominant somewhere. In order
to show how the dilaton and axion interaction with matter work, we
study the particular case $\tilde\lambda=0$ and let us
artificially drop out the matter interaction from the dilaton
equation (\ref{eq:phi}). In what follows we study this toy model.
%
%%%%%%%%%%%%%%%%%%%   Inicia parte de CQG %%%%%%%%%%%%%%%%%%
%
For this one it is convenient to change the variable $w$ for $
w=V_3$, with the definition of the potentials
$V_3=\sqrt{\kappa}\psi_0\psi$ such that
$V=V_\phi^2+1/4(V_1+V_2)^2\,V_3^2$. Thus, equations
(\ref{eq:x})-(\ref{eq:w}) transform into the new system
\begin{eqnarray}
&x^{\prime }& =-3\,x-\sqrt {\frac{3}{2}} \left(\lambda\,
uv+\frac{\alpha}{4}({v}+{u})^{2}{w}^{2} \right) +\frac{3}{2}
\Pi\,x, \nonumber
\\
&A^{\prime } &=-3\,A-\sqrt
{\frac{3}{2}}{\psi_0}\,\frac{1}{2}({v}+{u})^{2}{w}+\frac{3}{2}
\Pi\,A, \nonumber
\\
&y^{\prime } &=\frac{3}{2}\, \left( \Pi-1+
\alpha\sqrt{\frac{2}{3}}x\right) y, \nonumber
 \\
&z^{\prime } &=\frac{3}{2}\, \left( \Pi-\frac{4}{3}+
\alpha\sqrt{\frac{2}{3}}x \right)\,z, \nonumber
\\
&u^{\prime } &=\sqrt {\frac{3}{2}}\lambda\,v\,x+\frac{3}{2}
\Pi\,u, \nonumber
\\
&v^{\prime } &=\sqrt {\frac{3}{2}}\lambda\,u\,x+\frac{3}{2}
\Pi\,v, \nonumber
 \\
&l^{\prime } &=\frac{3}{2}\, \left( \Pi-{\gamma_{DE}} \right)\,l,
\nonumber
\\
 &w^{\prime } &=\sqrt {6}{\psi_0}\,A,
\label{eq:w_old}
\end{eqnarray}
The quantity $\Pi$ is now defined as
\begin{equation}
\Pi= 2\,{x}^{2}+2\,{A}^{2}+{y}^{2}+\frac{4}{3}\,{z}^{
2}+{\gamma_{DE}}\,{l}^{2}-\lambda\sqrt{\frac{2}{3}}
\left(y^2+z^2\right)x
\end{equation}
and the new Friedman constriction (\ref{eq:Friedman}) reads
\be x^2+A^2+y^2+z^2+u^2+l^2+\frac{1}{4}\,(u+v)^2\,w^2=1.
\label{eq:constriccion_old} \ee
The density quantities $\Omega_x$ now are
\begin{eqnarray}
\Omega_{DM}&=& x^2+u^2, \nonumber\\
\Omega_{DE}&=& l^2, \nonumber\\
\Omega_{b}&=& y^2, \nonumber\\
\Omega_{rad}&=& z^2, \nonumber\\
\Omega_{A}&=& 1-x^2-u^2-y^2-z^2-l^2, \label{eq:densidades_old}
\end{eqnarray}
where we have used the constriction (\ref{eq:constriccion_old}).
Equations (\ref{eq:w_old}) are now a new dynamical system.
%
%%%%%%%%%%%%%%%%%%%   Finaliza parte de CQG %%%%%%%%%%%%%%%%%%
%%
%
The evolution of this one is shown in Fig\ref{fig:pegado}. From
here we can see that now radiation dominates the early universe
without problems and that the behavior of the densities is again
very similar to the $\Lambda$CDM model but now for all redshifts.
The only difference is at redshifts $10^2<z<10^3$, where the
densities oscillate very hard. Unfortunately this time corresponds
to the dark age, when the universe has no stars and there is
nothing to observe which could give us some observational clue for
this behavior.

\begin{figure}[htb]
\begin{center}
\includegraphics[width=9cm]{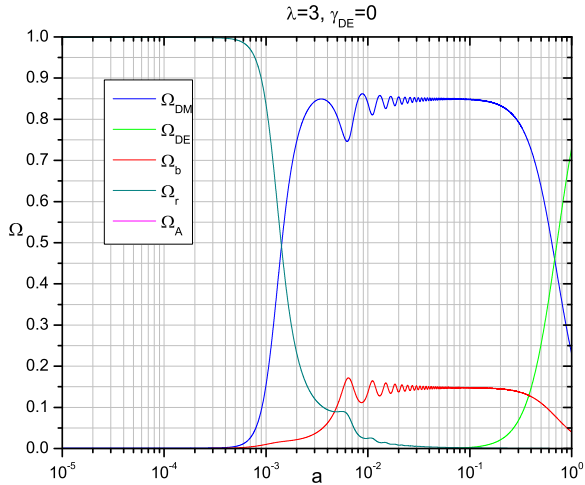}
\caption{\label{fig:cosh} Plot of the dynamics of the $\Omega$'s in
the cosh model. Here all the coupling constants of the superstrings
model $\alpha=\tilde\lambda=0$, $\lambda=3.0$, the initial values of
the dynamical variables at redshift $a=1$ are the same as in
Fig\ref{fig:completo1}. Observe how this theory predicts an
extremely similar behavior of the matter content of the universe as
the $\Lambda$CDM model, for all redshifts. Each curve contains over
$5\times 10^{5}$ points.}
\end{center}
\end{figure}

Finally, if the set both coupling constant
$\alpha=\tilde\lambda=0$, we recover a very similar behavior of
the densities to the $\Lambda$CDM model, this behavior es shown in
Fig.\ref{fig:cosh}. Observe here that the densities have not
oscillations any more, as in the $\Lambda$CDM model, supporting
the idea that it is just the coupling between dilaton, axion and
matter which makes difficult that the string theory reproduces the
observed universe.

\noindent {\it Structure Formation.-} As shown in figures
Fig.\ref{fig:completo1} and Fig.\ref{fig:pegado} the axion field
can be completely subdominant, but it can dominates the universe
at early times as in Fig.\ref{fig:completo2}. At late times,
$10^{-2}<a<1$, the structure formation is determined by the
dilaton field $\phi$ and its effective potential (\ref{eq:V}). In
\cite{Matos:2000ss} it was shown that the scalar field
fluctuations with a cosh potential follow the corresponding ones
of the cold dark matter (CDM) model for the linear regime. There,
it is shown that the field equations of the scalar field
fluctuations can be written in terms of the ones of the
$\Lambda$CDM model, in such a way that both models predict the
same spectrum in the linear regime of fluctuations.

\noindent{\it Galaxies Formation.-} Other main difference between
both models, the CDM and type IIB superstrings  is just in the
non-linear regime of fluctuations. Here numerical simulations show
that the scalar field virialize very early \cite{pacoluis},
causing that in the superstring model galaxies form earlier than
in the CDM paradigm. Furthermore, it has been shown that the
scalar field does not have a cuspy central density profile
\cite{nocuspCosh}. Numerical and semi-analytic simulations have
shown that the density profiles of oscillations (collapsed scalar
fields) are almost flat in the center
\cite{Matos:2003pe},\cite{Alcubierre:2001ea},\cite{pacoluis}. It
has been also possible to compare high and low surface brightness
galaxies with the scalar field model and the comparison shows that
there is a concordance between the model and the observations,
provided that the values of the parameters are just
$V_0\sim(3\times10^{-27}m_{Planck})^4$, $\lambda\sim20$
\cite{Matos:2003pe}. With this values of $V_0$ and $\lambda$, the
critical mass for collapse of the scalar field is just
$10^{12}M_{\odot}$ \cite{Alcubierre:2001ea}, as it is expected for
the halos of galaxies. These two features of the scalar field
collapse might give distinctive features to superstring theory. At
the present time there is a controversy about the density profiles
of the dark matter in the centers of the galaxies
\cite{cotrovCusp}. This model of the superstring theory predicts
that the center of the galaxies contains an almost flat central
density profile. We are aware that this result corresponds to the
particular compactification ${\bf T}^6/\mathbb{Z}_2$, but it could
be a general signature of string theory, in the sense that it
could survive in a more realistic compactification (including
branes and fluxes), that give rise to models that resemble the
Standard Model. In this case, if the cuspy dark matter density
profiles are observed or explained in some way, this model would
be ruled out. But if these profiles are not observed, it would be
an important astrophysical signature of string theory.

\section {Conclusions.}
In this work we propose an alternative interpretation of the
dilaton field in the type IIB supergravity on the ${\bf
T}^6/\mathbb{Z}_2$ orientifold model with fluxes \cite{Frey02}.
This alternative interpretation allowed us to compare this model
with the $\Lambda$CDM one, which has been very successful in its
predictions. The result is that, at lest in this model, radiation
seems to be subdominant everywhere, provoking difficulties to
explain big bang nucleosynthesis. Even when we see that in this
particular toy model, the behavior seems to be generic for all
strings theories. If this is the case, it is possible that the
dilaton and axion fields could not be able to be interpreted as
dark matter or dark energy, thus ether we should seek other
candidates and explain why we don't see the dilaton and axion
scalar fields in our observations, or we have to explain big bang
nucleosynthesis using the conditions given by superstrigs theory
we showed here, or we have to look for a mechanism to eliminate
the coupling between dilaton and axion with matter at very early
times. This last option is maybe more realistic. Even if we solve
the radiation dominance problem, there are some differences
between $\Lambda$CDM and superstrings theory between
$10^2<z<10^3$, because string theory predicts around 16 millon
years of densities oscillations during the dark age. Nevertheless,
both models are very similar at late times, between $0<z<10^2$,
maybe the only difference during this last period is their
predictions on substructure formation and galaxies centers. While
CDM predicts much more substructure in the universe and very sharp
density profiles, scalar fields predict few substructure and
almost constant density profiles in centers of galaxies. The
confirmation of this observations could decide between these two
models. We are aware that this is orientifold model is still a toy
model and it would be interesting to study more realistic
compactifications (including brane and orientifold configurations)
and see if our results, including that of the dark matter density
profiles, survive and become a general feature of string theory.
If this is the case, this alternative interpretation of the fields
of the theory might permit to establish a contact of string theory
with the astrophysics phenomenology of dark matter, $i.e.$, its a
contact with future astrophysical and cosmological observations.
We conclude that this interpretation can give us a closer
understanding of superstrings theory with cosmology.

---------------------------------------

%%%%%%%%%%%%%%%%%%%%%%%%%%%%
%%%   ACKNOWLEDGMENTS   %%%
%%%%%%%%%%%%%%%%%%%%%%%%%%%%

\section{Acknowledgments}
The authors wish to thank Luis Ure\~na-L\'opez for the deep
reading of the paper, Norma Quiroz and Cesar Terrero for many
helpful discussions and for pointing out to us the reference
\cite{Frey02}. Correspondence with S. Sethi and A. Krause is
appreciated. TM wants to thank Matt Choptuik for his kind
hospitality at the UBC. This work was partly supported by CONACyT
M\'exico, under grants 32138-E, 42748, 47641 and 45713-F.

%%%%%%%%%%%%%%%%%%%%%%
%%%   REFERENCES   %%%
%%%%%%%%%%%%%%%%%%%%%%

\end{document}